\documentclass[preprint,aps]{revtex4}
\begin{document}
\title{Black-Scholes-Like Derivative Pricing With 
Tsallis Non-extensive Statistics}
\author{Fredrick Michael and M.D. Johnson}
\email{mjohnson@ucf.edu}
\affiliation{Department of Physics, University of Central Florida, Orlando, FL
32816-2385}
\date{April 30, 2002}
\begin{abstract}
We recently showed that the S\&P500 stock market index is
well described by Tsallis non-extensive
statistics and nonlinear Fokker-Planck time evolution. 
We argued that these results should be applicable to a broad
range of markets and exchanges where anomalous diffusion and `heavy'
tails of the distribution are present. In the present work we examine
how the Black-Scholes derivative pricing formula is modified when
the underlying security obeys non-extensive statistics and Fokker-Planck
time evolution. We answer this by recourse to the
underlying microscopic Ito-Langevin stochastic differential equation 
of the non-extensive process. 
\end{abstract}
\maketitle

For certain stochastic systems there is an interesting connection between
statistics and dynamics. A family of nonlinear Fokker-Planck time-evolution
equations turns out to be solved by probability distributions which
are characterized by Tsallis non-extensive 
statistics\cite{tsallis,tsallis-bukman,zanette}. 
Recently we exploited this connection to
analyze the dynamics of the S\&P500 stock index\cite{mj1}, showing that
price-change distributions had both non-extensive form and Fokker-Planck
time evolution. We argued that the results should be applicable to the broad
range of markets and exchanges characterized by anomalous (super) diffusion
and `heavy' distribution tails\cite{econophysics,mantegna1,mantegna2}. In
this paper we now investigate how the Black-Scholes derivative pricing
formula\cite{black} is modified when the underlying security is described by
non-extensive statistics. This is based on an analysis of the microscopic
Ito-Langevin stochastic differential equation underlying the macroscopic
nonlinear Fokker-Planck equation\cite
{gardiner,plastino,borland,tsallis-bukman}.

\section{Non-extensive statistics and time evolution}

We begin by summarizing the probability distribution function (PDF) $P(S,t)$
which is obtained using non-extensive statistics. 
Denote the value of a security at a trading time $\tau$ by $\mbox{price}(\tau)$.
In the following we will take prices and times relative to the price at
some arbitrary fixed reference time $\tau_0$. Thus 
$S(t) = \mbox{price}(\tau_0+t)-\mbox{price}(\tau_0)$ 
is the (relative) security value at a (relative) trading time $t$.
The desired form of $P(S,t)$ is
obtained by maximizing an incomplete information-theoretic measure
equivalent to the Tsallis entropy: 
\begin{equation}
S_{q}=-{\frac{1}{1-q}}\left( 1-\int P(S,t)^{q}\,dS\right) ,
\end{equation}
subject to constraints on three moments\cite
{tsallis-bukman,zanette,mj1,borland,wang1}. The resulting PDF is 
\begin{equation}
P(S,t)=\frac{1}{Z(t)}\left\{ 1+\beta(t)(q-1)[S-\overline{S}(t)]^2
\right\}^{-\frac{1}{q-1}}.  \label{P}
\end{equation}
Here $q$ is a time-independent parameter indicating the degree of
non-extensivity or equivalently the incompleteness of the information
measure. $Z(t)$ is a normalization constant, $\overline{S}(t)$
is the mean, and $\beta(t)$ is related to the distribution's variance by
\begin{equation}
\sigma^{2}(t)=\int_{-\infty}^{\infty}[S-\overline{S}(t)]^{2}
P(S,t)dS=\left\{ 
\begin{array}{cl}
\frac{1}{(5-3q)\beta (t )}, & q<\frac{5}{3} \\ 
\infty , & q\geq \frac{5}{3}.
\end{array}
\right. 
\end{equation}

It was shown rather unexpectedly that distributions of this non-extensive
form turn out to solve a non-linear Fokker-Planck partial differential
equation \cite{tsallis-bukman,zanette,plastino} 
\begin{equation}
{\frac{\partial P(S,t)}{\partial t}}=-{\frac{\partial }{\partial S}}\left[
F(S)P(S,t)\right] +\frac{D}{2}{\frac{\partial ^{2}P(S,t)^{2-q}}{\partial
S^{2}}}.  
\label{fokker1}
\end{equation}
Here $F(S)=\mu S$ is a linear driving term dependent on the
market rate of return $\mu$.
Eq.~(\ref{fokker1}) is solved by distributions of the Tsallis form
Eq.~(\ref{P}) if
the parameters in the latter evolve in time according to
\begin{eqnarray}
\label{Sbar}
\overline{S}(t) &=& \overline{S}(t_1) e^{\mu(t-t_1)}\\
\label{beta}
\beta(t) &=& \Big\{\beta(t_1)^{-{3-q\over2}}e^{\mu(3-q)(t-t_1)}
\nonumber\\
&+&2D\mu^{-1}(2-q)
\left[ \beta(t_1)Z^2(t_1) \right]^{q-1\over2}
\left[ e^{\mu(3-q)(t-t_1)}-1 \right]\Big\}^{-\frac{2}{3-q}}\\
\label{Z}
Z(t)/Z(t_1) &=& [ \beta(t)/\beta(t_1) ]^{-\frac{1}{2}} .
\end{eqnarray}
Here $t_1$ is an arbitrary time; \textit{e.g.}, it could
be the shortest measured interval after the reference time $\tau_0$,
so that $t_1=\tau_1-\tau_0$ equals, say, one minute.

In \cite{mj1} we investigated price changes in the S\&P500 index. We showed
that price-change distributions were well-described by 
distributions of the non-extensive form Eq.~(\ref{P}) evolving
in time according to Eq.~(\ref{fokker1}).
The super-diffusion
and fat tails characterizing this market are both a consequence of a
non-extensivity parameter $q$ greater than unity. 

The nonlinear Fokker-Planck equation is a macroscopic description of how a
probability distribution evolves in time. It is connected to an Ito-Langevin
stochastic differential equation which describes how a particular trajectory
evolves\cite{borland,gardiner}. The Ito-Langevin equation can be written
in the general form
\begin{equation}
\frac{dS}{dt} = a(S,t) + b(S,t)\, \eta(t).
\label{ito1}
\end{equation}
with $a$ the drift coefficient and $b$ the diffusion coefficient. 
In the stochastic term $\eta(t)dt=dW(t)$ is the Wiener process \cite{gardiner}.
[$\eta (t)$ is a delta-correlated ($\left\langle \eta (t)\eta (t^{\prime
})\right\rangle =\delta (t-t^{\prime })$), normally-distributed noise with
unit variance ($\left\langle \eta (t)^{2}\right\rangle =1)$ and zero mean 
($\left\langle \eta (t)\right\rangle =0$).]

Eq.~(\ref{fokker1}) has a corresponding Ito-Langevin equation of the
form Eq.~(\ref{ito1}) with
\begin{equation}
a(S,t) = F(S) = \mu S,\quad b(S,t)=\sqrt{DP(S,t)^{1-q}}.
\label{ab}
\end{equation}
Here the Fokker-Planck equation's driving term $F(S)=\mu S$ appears as a
time-independent linear drift coefficient $a$. 
The diffusion coefficient in 
our case is $b=\sqrt{DP(S,t)^{1-q}}$, which exhibits explicitly at
the level of the microscopic stochastic process the statistical dependence
of subsequent price changes on the macroscopic PDF $P(S,t)$. That is, the
memory effect representing correlations in time enters here simply via the
diffusion coefficient. We argued in \cite{mj1} that nonlinear
Fokker-Planck time evolution can be expected in any stochastic system in
which memory effects can be approximated in this simple manner as a
probability-dependent diffusion coefficient.

\section{Derivative pricing for non-extensive statistics}

Now we turn to the question of how the Black-Scholes derivative pricing
model is to be modified when the underlying security has non-extensive
statistics and nonlinear Fokker-Planck dynamics. 

We can define one form of portfolio
$\Pi =-G+S\frac{\partial G}{\partial S}$ \cite{black,merton1,cox1}.
This is short one share of a derivative $G$ and long 
$\partial G/\partial S$ shares of the underlying security (stock, say) $S$. 
The change in the value of the portfolio in a time $dt $ is 
\begin{equation}
d\Pi =-dG+dS\frac{\partial G}{\partial S}.  \label{dpi}
\end{equation}
(The number of stock shares $\partial G/\partial S$ is of course
constant during $dt $.) The change $dG$ during $dt $ is given by Ito's
formula \cite{gardiner}. This results from Taylor expanding to first order
in $dt $ and to second in order in price change $dS$,
and using $dW(t)^2=dt$:
\begin{eqnarray}
dG &=&\frac{\partial G}{\partial t }dt +\frac{\partial G}{\partial S}
dS+\frac{1}{2}\frac{\partial ^{2}G}{\partial S^{2}}dS^{2}  \nonumber \\
&=&\left( \frac{\partial G}{\partial t }+{\frac{1}{2}}b^2(S,t)
\frac{\partial ^{2}G}{\partial S^{2}}\right) dt +
\frac{\partial G}{\partial S}
dS.  \label{taylor1}
\end{eqnarray}
Using Eq.~(\ref{ab}), this becomes the stochastic differential equation
obeyed by the derivative $G(S,t )$.

We seek a portfolio which instantaneously earns the same rate of return $r$
as a short term risk-free security (assuming no arbitrage). Then 
\begin{equation}
d\Pi =r\Pi dt =r\left( -G+S\frac{\partial G}{\partial S}\right) dt .
\label{free}
\end{equation}
Substituting Eq.~(\ref{taylor1}) into Eq.~(\ref{dpi}) and equating the
result to Eq.~(\ref{free}) gives
\begin{equation}
\frac{\partial G(S,t )}{\partial t }+rS\frac{\partial G(S,t
)}{\partial S}+\frac{1}{2}b^2(S,t )\frac{\partial
^{2}G(S,t )}{\partial S^{2}}=rG(S,t ).  \label{black15}
\end{equation}
This pricing equation is analogous to the Black-Scholes result \cite{black},
generalized for an arbitrary diffusion term $b(s,t)$ \cite{cox1}.
The explicit dependence on the market rate of return $\mu$ has been
replaced by the risk-free rate $r$.

There is however a difficulty hidden in Eq.~(\ref{black15}). For our
case the diffusion term $b(S,t)=\sqrt{DP^{1-q}}$
depends on the probability distribution function of the underlying
stock. Hence Eq.~(\ref{black15}) depends \textit{implicitly} on the market's
rate of return $\mu$. To show what difficulty this entails, let us
begin by reviewing the Cox and Ross approach to solving Eq.~(\ref{black15}).

First define a two-point function $P(S,t|S',t')$ which obeys an equation
very similar to Eq.~(\ref{fokker1}),
\begin{equation}
{\frac{\partial P(S,t|S',t')}{\partial t}}=-{\frac{\partial }{\partial S}}\left[
\mu S P(S,t|S',t')\right] +\frac{1}{2}
{\frac{\partial ^{2}}{\partial S^{2}}}  
\left[ b^2(S,t) P(S,t|S',t') \right],
\label{fokker2}
\end{equation}
but with a boundary condition $P(S,t'|S',t')=\delta(S-S')$. 
[Here we continue to use $b=\sqrt{DP(S,t)^{1-q}}$.]
The Cox-Ross solution is based on the fact that Eq.~(\ref{fokker2}) is a forward
Chapman-Kolmogorov equation, and as such also has a corresponding
backwards form \cite{gardiner,cox1}:
\begin{equation}
{\frac{\partial P(S,t |S^{\prime },t ^{\prime })}{\partial t
^{\prime }}}=-\mu S^{\prime} 
{\frac{\partial P(S,t |S^{\prime},t^{\prime})}{\partial S^{\prime }}}
-\frac{1}{2}b^{2}(S',t')
{\frac{\partial ^{2}P(S,t
|S^{\prime},t^{\prime})}{\partial S^{\prime }{}^{2}}}.
\label{backwards}
\end{equation}
As an example, let us seek a solution of Eq.~(\ref{black15}) for a
European-style call option $G(S,t)$ on a non-dividend-paying stock $S$.
Following Cox and Ross, we try the form
\begin{equation}
G(S,t)=e^{-r(T-t )}\int G(S_T,T) \tilde{P}(S_T,T|S,t)\,dS_T.
\label{G}
\end{equation}
This involves the value of $G$ at the maturity time $T$. For the European
call option this is
\begin{equation}
G(S_T,T)=\max(S_T-X,0),
\end{equation}
where $S_T$ is the terminal stock price and $X$ the exercise price.
(At maturity the value of the call option is worthless if the terminal
stock price is less than the exercise price; otherwise the value is
the price difference.) Substituting Eq.~(\ref{G}) into Eq.~(\ref{black15}),
one finds that $\tilde{P}$ must solve
\begin{equation}
{\frac{\partial \tilde{P}(S_T,T |S,t)}{\partial t}}
=-r S {\frac{\partial \tilde{P}(S_T,T|S,t)}{\partial S}}
-\frac{1}{2}b^{2}(S,t) 
{\frac{\partial ^{2}\tilde{P}(S_T,T |S,t)}{\partial S^2}},
\label{backwards1}
\end{equation}
with boundary condition $\tilde{P}(S_T,T|S,T)=\delta(S_T-S)$.
This has the form of the backwards Chapman-Kolmogorov equation
(\ref{backwards}) but with $\mu$ replaced by the risk-free rate $r$.

In all the cases considered by Cox and Ross, the diffusion term
$b^2$ was independent of the underlying stock's rate of return $\mu$.
Then $\tilde{P}$ was the probability distribution for a risk neutral
world, and could readily be found.

Here however the diffusion term $b^2$ depends implicitly on the
underlying stock's rate of return $\mu$,
and consequently Eq.~(\ref{backwards1}) cannot
be solved by assuming a risk neutral world. Assuming risk neutrality
amounts to replacing $\mu$ by the risk-free rate $r$ everywhere.
Then Eq.~(\ref{backwards1}) would be identical to Eq.~(\ref{backwards})
with $\mu$ replaced by $r$ everywhere. This could be solved by
a Tsallis form. Unfortunately this replacement is not justified,
and we have to turn to an alternative approach.

In fact, we have found two variations on the Cox-Ross approach which
permit straightforward solutions of the valuation equation for
securities with non-extensive statistics. We will present both.

The first amounts to a change of variables. Define $\tilde{S}=\tilde{S}(S,t)$
as some function of $S,t$. Then using Ito's formula as in Eq.~(\ref{taylor1}),
we have
\begin{equation}
d\tilde{S} = \tilde{a}\, dt + \tilde{b}\, dW
\end{equation}
where
\begin{equation}
\tilde{a}(\tilde{S},t) = 
\frac{\partial \tilde{S}}{\partial t} + \frac{1}{2}b^2
\frac{\partial^2 \tilde{S}}{\partial S^2} +
a\frac{\partial \tilde{S}}{\partial S}, \qquad
\tilde{b}(\tilde{S},t) = b\frac{\partial \tilde{S}}{\partial S}.
\label{b2}
\end{equation}
Options written on $\tilde{S}$ can be evaluated for any $\tilde{b}$
which is independent of the original security's market rate of return $\mu$.
Let us consider the simplest case, where $\tilde{b}=\tilde{b}(t)$ is
a function of $t$ only. Then the second of Eqs.~(\ref{b2}) can be solved for
$\tilde{S}$:
\begin{equation}
\tilde{S}(S,t) = \frac{\tilde{b}(t)}{\sqrt{D}Z^{\frac{q-1}{2}}}
\frac{ \sinh^{-1} \left[ \sqrt{\beta(q-1)} (S-\overline{S}) \right]}
{\sqrt{\beta(q-1)}}.
\end{equation}
Here we have used Eqs.~(\ref{ab},\ref{P}).

Now we consider an option $G(\tilde{S},t)$ and a portfolio
$\Pi = -G + \frac{\partial G}{\partial \tilde{S}}\tilde{S}$.
An analysis exactly like that leading to Eq.~(\ref{black15}) shows
that $\Pi$ follows the risk-free rate of return $r$ if
\begin{equation}
\frac{\partial G}{\partial t }+
r\tilde{S}\frac{\partial G}{\partial \tilde{S}}+
\frac{1}{2}\tilde{b}^2
\frac{\partial^{2}G}{\partial \tilde{S}^{2}}=rG.  \label{black16}
\end{equation}
This by construction has no dependence on $\mu$, and hence describes
a risk-free universe as in Cox and Ross \cite{cox1}. In practical applications,
$\tilde{S}$ can be viewed as a derivative of the original non-extensive
security $S$, and $G(\tilde{S},t)$ is then an option involving $\tilde{S}$.

A very different route to valuing options for non-extensive securities
comes from converting the market's Ito-Langevin equation into a coupled
process with a constant diffusion coefficient. This uses an idea developed
for time nonhomogeneous systems \cite{gardiner}. Consider the coupled process
\begin{eqnarray}
\label{hatS}
d \hat{S} &=& \left[ \mu S(t) + b(S,t) y(t) \right] dt
\equiv \hat{a}\,dt \\
d y &=& -\gamma\, y(t) dt + \gamma\, dW(t) ,
\label{y}
\end{eqnarray}
where $\gamma$ is a constant. Notice that $\hat S$ has no diffusion
term; this will permit us to solve a two-variable Black-Scholes-like
equation (below). First, however, we need to relate the coupled process to the
original security $S$.

Eq.~(\ref{y}) is formally solved by
\begin{equation}
y(t) = \gamma \int_{-\infty}^t e^{-\gamma(t-t')} \eta(t') dt'.
\end{equation}
In the limit $\gamma\rightarrow\infty$ this
becomes a stationary, $\delta$-correlated Gaussian process \cite{gardiner}.
That is,
\begin{equation}
y(t) \rightarrow \eta(t) \mbox{\ as\ }
{\gamma\rightarrow\infty} .
\end{equation}
Consequently as $\eta\rightarrow\infty$ Eq.~(\ref{hatS}) becomes identical
to our market Ito-Langevin equation [Eq.~(\ref{ito1})] and hence
$\hat S$ becomes the original security $S$. We can then analyze
Eqs.~(\ref{hatS},\ref{y}) for finite $\gamma$ and take
$\gamma\rightarrow\infty$ at the end.

Consider an option $G(\hat{S},y,t)$. For finite $\gamma$, 
Eq.~(\ref{hatS}) has no diffusion term. Consequently to find $dG$
we expand to first order in $t$ and $\hat S$, and second order in $y$.
The result is
\begin{equation}
dG = \left(
\frac{\partial G}{\partial t} + \hat{a}\frac{\partial G}{\partial \hat{S}}
+ \frac{\gamma^2}{2} \frac{\partial^2 G}{\partial y^2} \right) dt
+ \frac{\partial G}{\partial y}dy .
\end{equation}
Now construct a portfolio 
$\Pi = -G +\frac{\partial G}{\partial \hat{S}}\hat{S} + \frac{\partial G}{\partial y}y$.
One readily finds that $\Pi$ evolves at the risk-free rate $r$ if
\begin{equation}
\frac{\partial G}{\partial t }+
r\hat{S}\frac{\partial G}{\partial \hat{S}}+
ry \frac{\partial G}{\partial y} +
\frac{\gamma^2}{2}
\frac{\partial^{2}G}{\partial y^2}=rG.  \label{black17}
\end{equation}
We see that again the market rate of return $\mu$ has dropped out.
Thus we have obtained a two-variable form of the usual Black-Scholes
equation. Solutions can be find exactly as in Cox and Ross \cite{cox1}:
\begin{equation}
G(\hat{S},y,t) = e^{-r(T-t )}\int G(S_T,y_T,T) \hat{P}(S_T,y_T,T|S,y,t)\,dS_T\,dy,
\end{equation}
where $\hat{P}$ solves the backwards equation
\begin{equation}
{\frac{\partial \hat{P}(\hat{S}_T,y_T,T |\hat{S},y,t)}{\partial t}}
= -r \hat{S} {\frac{\partial \hat{P}}{\partial \hat{S}}}
 -r y {\frac{\partial \hat{P}}{\partial y}}
-\frac{\gamma^2}{2} 
{\frac{\partial^2\hat{P}}{\partial y^2}}.
\label{backwards2}
\end{equation}
Practical application of this method of valuation would require solving with
a finite value of $\gamma$, but a value large enough so that $y(t)$ is
sufficiently close to $\eta(t)$ for time scales considered.

We can make some connection between the two approaches described above
by integrating out $y$. One can define
\begin{equation}
G(\hat{S},t) = \int dy\, G(\hat{S},y,t)
= \int d\hat{S}_T \,G(\hat{S}_T,T) \hat{P}(\hat{S}_T,T|\hat{S},t),
\label{intG}
\end{equation}
where
\begin{eqnarray}
\hat{P}(\hat{S}_T,T|\hat{S},t) &=& \frac{1}{G(S_T,T)}
\int dy \,dy_T \,G(\hat{S}_T,y_T,T) \hat{P}(\hat{S}_T,y_T,T|\hat{S},y,t)\\
G(\hat{S}_T,T) &=& \int dy_T \,G(\hat{S}_T,y_T,T).
\end{eqnarray}
Eq.~(\ref{intG}) is formally equivalent to the solution of
Eq.~(\ref{black16}) in the Cox-Ross form Eq.~(\ref{G}), in the limit
$\gamma\rightarrow\infty$.

We have followed the lines of Black-Scholes and Cox-Ross to develop
an options pricing formula for securities obeying non-extensive statistics
and nonlinear Fokker-Planck time evolution. We showed in \cite{mj1} that the
S\&P500 index is well described using this approach, and argued there that a
description in terms of non-extensive statistics can be useful for any
market in which the stylized facts of fat tails and anomalous diffusion are
pronounced. The pricing formula obtained here would then be useful for
options based on any such market. As an example we obtained pricing models
for a European style call option on an underlying asset that pays out no
dividends. 

We acknowledge support from the NSF through grant DMR99-72683.

\end{document}